\documentclass[aip,twocolumn,superscriptaddress,showpacs]{revtex4}
\usepackage{graphicx}
\usepackage{bm}
\usepackage{color}
\usepackage{dcolumn}
\usepackage{tabularx}
\usepackage{textcomp}

\begin{document}
\title{Proton radiography of magnetic field produced by ultra-intense laser irradiating capacity-coil target}
\author{W. W. Wang}
\affiliation{Science and Technology on Plasma Physics Laboratory, Research Center of Laser Fusion, China Academy of Engineering Physics (CAEP), Mianyang 621900, China}
\author{J. Teng}
\affiliation{Science and Technology on Plasma Physics Laboratory, Research Center of Laser Fusion, China Academy of Engineering Physics (CAEP), Mianyang 621900, China}
\author{J. Chen}
\affiliation{Science and Technology on Plasma Physics Laboratory, Research Center of Laser Fusion, China Academy of Engineering Physics (CAEP), Mianyang 621900, China}
\author{H. B. Cai}
\affiliation{Institute of Applied Physics and Computational Mathematics, Beijing 100094, China}
\affiliation{Center for Applied Physics and Technology, Peking University, Beijing 100871, China}
\author{S. K. He}
\affiliation{Science and Technology on Plasma Physics Laboratory, Research Center of Laser Fusion, China Academy of Engineering Physics (CAEP), Mianyang 621900, China}
\author{W. M. Zhou}
\affiliation{Science and Technology on Plasma Physics Laboratory, Research Center of Laser Fusion, China Academy of Engineering Physics (CAEP), Mianyang 621900, China}
\author{L. Q. Shan}
\affiliation{Science and Technology on Plasma Physics Laboratory, Research Center of Laser Fusion, China Academy of Engineering Physics (CAEP), Mianyang 621900, China}
\author{F. Lu}
\affiliation{Science and Technology on Plasma Physics Laboratory, Research Center of Laser Fusion, China Academy of Engineering Physics (CAEP), Mianyang 621900, China}
\author{Y. C. Wu}
\affiliation{Science and Technology on Plasma Physics Laboratory, Research Center of Laser Fusion, China Academy of Engineering Physics (CAEP), Mianyang 621900, China}
\author{W. Hong}
\affiliation{Science and Technology on Plasma Physics Laboratory, Research Center of Laser Fusion, China Academy of Engineering Physics (CAEP), Mianyang 621900, China}
\author{B. Bi}
\affiliation{Science and Technology on Plasma Physics Laboratory, Research Center of Laser Fusion, China Academy of Engineering Physics (CAEP), Mianyang 621900, China}
\author{F. Zhang}
\affiliation{Science and Technology on Plasma Physics Laboratory, Research Center of Laser Fusion, China Academy of Engineering Physics (CAEP), Mianyang 621900, China}
\author{D. X. Liu}
\affiliation{Science and Technology on Plasma Physics Laboratory, Research Center of Laser Fusion, China Academy of Engineering Physics (CAEP), Mianyang 621900, China}
\author{F. B. Xue}
\affiliation{Institute of Nuclear Physics and Chemistry, CAEP, Mianyang 621900, China}
\author{B. Y. Li}
\affiliation{Science and Technology on Plasma Physics Laboratory, Research Center of Laser Fusion, China Academy of Engineering Physics (CAEP), Mianyang 621900, China}
\author{B. Zhang}
\affiliation{Science and Technology on Plasma Physics Laboratory, Research Center of Laser Fusion, China Academy of Engineering Physics (CAEP), Mianyang 621900, China}
\author{Y. L. He}
\affiliation{Science and Technology on Plasma Physics Laboratory, Research Center of Laser Fusion, China Academy of Engineering Physics (CAEP), Mianyang 621900, China}
\author{W. He}
\affiliation{Science and Technology on Plasma Physics Laboratory, Research Center of Laser Fusion, China Academy of Engineering Physics (CAEP), Mianyang 621900, China}
\author{J. L. Jiao}
\affiliation{Science and Technology on Plasma Physics Laboratory, Research Center of Laser Fusion, China Academy of Engineering Physics (CAEP), Mianyang 621900, China}
\author{K. G. Dong}
\affiliation{Science and Technology on Plasma Physics Laboratory, Research Center of Laser Fusion, China Academy of Engineering Physics (CAEP), Mianyang 621900, China}
\author{F. Q. Zhang}
\affiliation{Science and Technology on Plasma Physics Laboratory, Research Center of Laser Fusion, China Academy of Engineering Physics (CAEP), Mianyang 621900, China}
\author{Z. G. Deng}
\affiliation{Science and Technology on Plasma Physics Laboratory, Research Center of Laser Fusion, China Academy of Engineering Physics (CAEP), Mianyang 621900, China}
\author{Z. M. Zhang}
\affiliation{Science and Technology on Plasma Physics Laboratory, Research Center of Laser Fusion, China Academy of Engineering Physics (CAEP), Mianyang 621900, China}
\author{B. Cui}
\affiliation{Science and Technology on Plasma Physics Laboratory, Research Center of Laser Fusion, China Academy of Engineering Physics (CAEP), Mianyang 621900, China}
\author{D. Han}
\affiliation{Science and Technology on Plasma Physics Laboratory, Research Center of Laser Fusion, China Academy of Engineering Physics (CAEP), Mianyang 621900, China}
\author{K. N. Zhou}
\affiliation{Science and Technology on Plasma Physics Laboratory, Research Center of Laser Fusion, China Academy of Engineering Physics (CAEP), Mianyang 621900, China}
\author{X. D. Wang}
\affiliation{Science and Technology on Plasma Physics Laboratory, Research Center of Laser Fusion, China Academy of Engineering Physics (CAEP), Mianyang 621900, China}
\author{Z. Q. Zhao}
\affiliation{Science and Technology on Plasma Physics Laboratory, Research Center of Laser Fusion, China Academy of Engineering Physics (CAEP), Mianyang 621900, China}
\author{L. F. Cao}
\affiliation{Science and Technology on Plasma Physics Laboratory, Research Center of Laser Fusion, China Academy of Engineering Physics (CAEP), Mianyang 621900, China}
\author{B. H. Zhang}
\affiliation{Science and Technology on Plasma Physics Laboratory, Research Center of Laser Fusion, China Academy of Engineering Physics (CAEP), Mianyang 621900, China}
\author{X. T. He}
\affiliation{Institute of Applied Physics and Computational Mathematics, Beijing 100094, China}
\affiliation{Center for Applied Physics and Technology, Peking University, Beijing 100871, China}
\author{Y. Q. Gu}
\email{yqgu@caep.ac.cn}
\affiliation{Science and Technology on Plasma Physics Laboratory, Research Center of Laser Fusion, China Academy of Engineering Physics (CAEP), Mianyang 621900, China}
\affiliation{Center for Applied Physics and Technology, Peking University, Beijing 100871, China}
\date{\today}

\begin{abstract}
Ultra-intense ultra-short laser is firstly used to irradiate the capacity-coil target to generate magnetic field. The spatial structure and temporal evolution of huge magnetic fields were studied with time-gated proton radiography method. A magnetic flux density of 40T was measured by comparing the proton deflection and particle track simulations. Although the laser pulse duration is only 30fs, the generated magnetic field can last for over 100 picoseconds. The energy conversion efficiency from laser to magnetic field can reach as high as $\sim$20\%. The results indicate that tens of tesla(T) magnetic field could be produced in many ultra intense laser facilities around the world, and higher magnetic field could be produced by picosecond lasers.
\end{abstract}


\pacs{52.38.Fz, 52.65.Cc, 52.70.Ds, 52.57.Kk}
\maketitle
Laboratory production of large magnetic field is of interest to many research areas including atomic and molecular physics \cite{Gilch}, astrophysics \cite{Astrophys}, and inertial confinement fusion research \cite{ICF}. A particular example is the collimation of fast electron beam, which is an important issue in the Fast Ignition (FI) scheme \cite{FI}. Using external magnetic field produced by laser driving capacity-coil target \cite{Daido,Courtois,Fujioka2013} is one of the promising approaches \cite{Strozzi,Cai2013,GuYQ}.

In the FI scheme, the deuterium-tritium pellet is firstly compressed to high density. When the pellet is close to maximum compression, a portion of it is heated by an intense flux of electron beam generated in the ultra-intense laser-plasma interactions \cite{Tabak}. Recent experiments and simulations \cite{Green} have shown that the divergence of fast electron beam can be as high as 50 degrees, which reduces the energy coupling efficiency from fast electrons to hot-spot greatly \cite{Solodov}. Strozzi $et\ al.$ \cite{Strozzi} and Cai $et\ al.$ \cite{Cai2013} show that external axial magnetic field could collimate the electron beam efficiently, especially for 300-3000T magnetic field \cite{Cai2013}. Fujioka $et\ al.$ \cite{Fujioka2013} have demonstrated magnetic field above 1kT produced by nanosecond laser irradiating capacity-coil target, which shows an increasing B-field with the laser intensity. However, no experimental results at relativistic laser intensity was given. Another question is that the total energy of the magnetic field generated in Ref.\ [\onlinecite{Fujioka2013}] is several times larger than the laser energy. The problem may come from the detectors. The traditional detector for magnetic field produced by laser driving capacity-coil target is B-dot probe, which is an induction coil positioned around the laser-driven coil target. But the signal is easily affected by electric and magnetic noise generated by laser-plasma interactions. Fujioka $et\ al.$ \cite{Fujioka2013} used faraday rotation method to detect the magnetic field in the experiment, but the magnetic signal was blacking out when it increased. As a result, new probing method should be developed for magnetic field generated by laser irradiating capacity-coil target.

In this paper, ultra-intense ultra-short laser was firstly used to generate magnetic field by irradiating capacity-coil target. And proton radiography \cite{Kugland,Willingale} was used to detect the magnetic field, which has high temporal and spatial resolution. The spatial structure and temporal evolution of huge magnetic fields were observed in experiments. It is found that the magnetic field reaches the largest magnitude at 50-60ps when ultra-intense ultra-short laser is used. The time scale cannot be differentiated by the traditional ways (the B-dot Probe or Faraday rotation method). Although the laser pulse duration is only 30fs, the generated magnetic field can last for over 100ps due to the increasing of resistance in U-turn coil. The energy couple efficiency from laser to magnetic field is as high as $\sim$20\% in our experiments. Two pick-up coil probes are used synchronously to detect the magnetic field. But they are greatly affected by the electromagnetic noise around the target chamber, and their results are not presented in this paper.

\begin{figure}
\includegraphics[width=0.7\columnwidth]{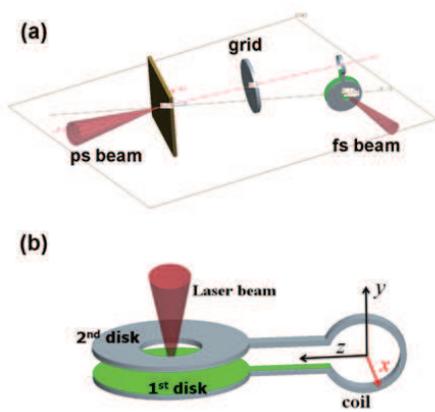}
\caption{(color online) (a) Experiment setup: picosecond laser beam irradiates Cu foil target and femtosecond laser beam irradiates capacity-coil target, and the mesh grid is positioned in the middle of them. (b) The detail of the capacity-coil target, which is made of nickel(Ni).\label{FIG.set}}
\end{figure}

The experiment was carried out using the XingGuang-\uppercase\expandafter{\romannumeral3} facility at Science and Technology on Plasma Physics Laboratory in Research Center of Laser Fusion, CAEP. The set up of the experiments is illustrated schematically in Fig.\ \ref{FIG.set}(a). The magnetic field was generated by capacity-coil target driven by femtosecond laser beam, whose wavelength was 800 nm with $\sim$30fs duration. The laser energy was 9$\pm$2J and laser intensity was above $10^{18}\ {\rm W/cm^2}$ in experiments. It irradiated on 80 $\mu$m Ni piece of the capacity-coil target (the 1st disk in Fig.\ \ref{FIG.set}(b)) with an incident angle of $21^{\circ}$. The proton beam was produced by picosecond laser irradiating 5 $\mu$m Cu foil with $24^{\circ}$ incident angle, producing protons by TNSA (Target Normal Sheath Acceleration) mechanism \cite{TNSA} with energy up to 13 MeV. The laser wavelength was 1.06 $\mu$m with $\sim$1.25ps duration. Its energy is 100$\pm$20J and the intensity was about $1\times10^{19}\ {\rm W/cm^2}$. The proton beam was perpendicular to the coil plane, as shown in Fig.\ \ref{FIG.set}(a).

The capacity-coil target is shown in Fig.\ \ref{FIG.set}(b). Two nickel(Ni) disks are connected by a U-turn coil. The thickness of the capacitance disks is 80 $\mu$m, and the cross section of the coil line is $80\ \mu m\times200\ \mu m$ rectangle. The diameter of the circle part of U-turn coil is 1000 $\mu$m, and the length of the straight lines are both 1000 $\mu$m, which was designed to avoid the laser plasma interaction noise in the 1st disk. The detector is a stack of radiochromic film (RCF) where the protons will deposit most of their energy at a depth within the stack according to their Bragg curve \cite{Willingale}. The stack is composed of several pieces of Gafchromic HD-810 film with a 20 $\mu$m Al foil filter at the very front to block unwanted radiation. A Ni mesh grid was placed 5 mm from the Cu foil target. The distance from the Cu foil to U-turn coil plane is 10 mm, and to RCF is 100 mm.


The mechanism of magnetic field production in the paper is somewhat different from the ns laser beam driven case. Courtois $et\ al.$ \cite{Courtois} have described the latter case in detail. In our experiments, ultra intense laser can produce $\sim$MeV electrons in front and rear sides of the 1st disk on capacity-coil target. Electrons escaping from both sides are contributing to the electrical potential between the capacitance disks. The duration of driving laser is very short ($\sim$30fs), which is much smaller than the generation time of magnetic field. Thus the magnetic field increasing process can be treated as LRC circuit, whose initial voltage in the capacitance is created by the $\sim$MeV electrons. It is estimated that the U-turn coil inductance $L=2.45\times10^{-9}$ H and capacitance $C=6.6\times10^{-14}$ F for the target used in experiments. The magnetic field increasing time is $2\pi \sqrt{CL}/4\simeq$20ps (the coil resistance in room temperature is too small to be neglected). When thermal electrons with lower energy are considered, the time when magnetic field reaches maximum would be longer than 20ps.

\begin{figure}
\includegraphics[width=0.8\columnwidth]{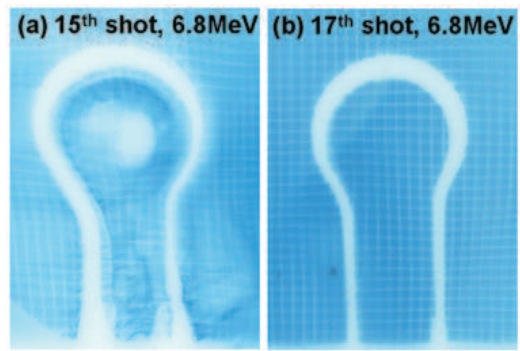}
\caption{(color online) The RCF images from proton radiography in experiments are presented. (a) Femtosecond laser beam was used to produce magnetic field. (b) No femtosecond laser beam existed, and other experimental setups were the same as in (a). \label{FIG.experiment}}
\end{figure}

The RCF images from proton radiography in experiments are shown in Fig.\ \ref{FIG.experiment}. The deflections of protons around the U-turn coil and in the center of circle part are observed when magnetic field is generated by femtosecond laser beam, as shown in Fig.\ \ref{FIG.experiment}(a). On the other hand, the deflection of protons almost disappears when the femtosecond laser beam is not used to produce the magnetic field, as shown in Fig.\ \ref{FIG.experiment}(b). Therefore the electromagnetic noise comes from the picosecond laser beam (producing protons) can be neglected, which makes sure that the deflection is caused by the magnetic field produced by femtosecond laser beam.

\begin{figure}
\includegraphics[width=0.9\columnwidth]{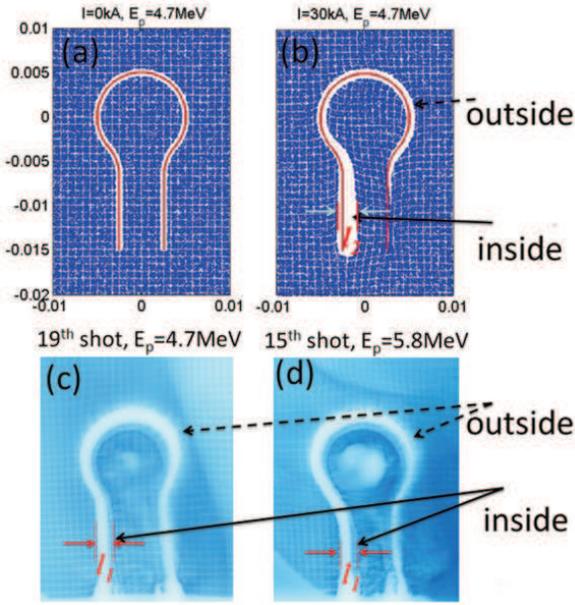}
\caption{(color online) Particle track simulations (a,b) and exprimental results (c,d) of proton images. (a) current in U-turn coil $I=0$, proton energy $E_p=4.7$ MeV; (b) $I=20$ kA, $E_p=4.7$ MeV; (c) 3rd piece of RCF image in 19th shot, i.e. $E_p=4.7$ MeV; (d) 15th shot, $E_p=5.8$ MeV. $l_1$ and $l_2$ in figures are the distance of proton deflection in straight line part.\label{FIG.simulation}}
\end{figure}

Particle track simulations are performed to calculate the magnetic field. We suppose the magnetic field only comes from the current in the U-turn coil. The magnitude of magnetic field is determined by the current, and its distribution is determined by the shape of the coil. Protons are running in the 3-dimensional magnetic field to get the simulation results. The space-charge effects in proton beam are not considered in particle track simulations. The proton deflections around the U-turn coil in simulations are in accordance with the experimental results under this assumption, as shown in Fig.\ \ref{FIG.simulation}. But the proton deflections in the center of the circle part cannot be observed in the simulations. It is caused by the space charge effects between the electrons and protons in the probe beam (proton beam), which cannot be simulated by the particle track simulation. The 2D3V Particle-in-Cell (PIC) simulations \cite{ZhangZM} are used to explain it, where the space charge effects are taken into account.

\begin{figure}
\includegraphics[width=1.0\columnwidth]{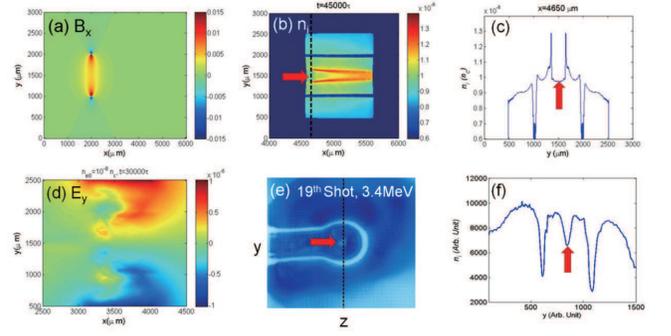}
\caption{(color online) 2D3V PIC simulations (a-d) and exprimental results (e,f). (a) The external magnetic field in $x$ direction ($B_x$) in simulations, and the simulation plane is the $x-y$ plane with $z=0$ in Fig.\ref{FIG.set}(b). (b) The density distribution of protons at time $t=45000\tau$ and (c) is a special case at position $x=4650\mu m$. (d) The electrical field in $y$ direction ($E_y$) at time $t=35000\tau$. (e) One RCF image in experiment, and the hollow structure of protons is in the center of the circle part. (f) is a special case of Fig.(e) at position around $z=0$.\label{FIG.pic}}
\end{figure}

As is known, when protons are accelerated, almost the same number of electrons are in the beam to keep the proton beam neutral. Therefore, a plasma beam formed by electrons and protons with the same number is set in PIC simulations. The simulation is performed in the $x-y$ plane with $z=0$ in Fig.\ \ref{FIG.set}(b). The external magnetic field is added, which produced by 40kA current (corresponding to the maximum magnetic field in our experiment) in circle coil, as shown in Fig.\ \ref{FIG.pic}(a). The simulation grid size is $\Delta x=\Delta y=1\ \mu$m with $6000\times3000$ grid cells, and the time step used is $\tau =$3.3fs. The initial position of plasma beam is in box of 100 $\mu$m$<x<$1100 $\mu$m and 500 $\mu$m$<y<$2500 $\mu$m, with velocity of 0.1$c$ in $x$ direction (where c is the light speed, and 0.1$c$ is the speed of 4.7 MeV proton beam). The density of the plasma beam is $10^{-8}n_c$ ($n_c=10^{21}$ cm$^{-3}$), which estimated from 4.7 MeV protons with energy 0.1J and space size $4\times4\times1$ mm$^3$. Fig.\ \ref{FIG.pic}(b) shows the proton density distribution at simulation time of $t=45000\ \tau$. The structure is in accordance with the experimental results on RCF, especially the hollow structure is also formed in the center, as compared in Fig.\ \ref{FIG.pic}(c) and (f). The physical model is described below. As the electron mass is very small, electrons will be focused by the magnetic field when they are close to the target coil, and then defocused when away from the coil. The electron rotation radius $R_c$ around coil can be estimated as $R_c=m_ev/(Bq)$, where $m_e$ is the electron mass, $q$ is the electron charge, and $v$ is the velocity. When magnetic field $B=10$T (which can be obtained in our experiments), the $R_c=17$ $\mu$m, which is much smaller than the coil Radius. Thus electrons running with proton beam will be focused and then defocused by the magnetic field. Thus besides the magnetic field, protons will be deflected by the electrical potential formed from the deviation of electrons and protons. The electrons focusing and defocusing are observed in PIC simulations (not shown), and the electrical field with magnitude of $10^7$ V/m who excluding the protons from center to outside is shown in Fig.\ref{FIG.pic}(d). The results of PIC simulations indicate that electrons in proton beam may have great influence on the proton deflections in Proton Radiography. It is necessary to shield the electrons in the proton beam.

\begin{figure}
\includegraphics[width=0.9\columnwidth]{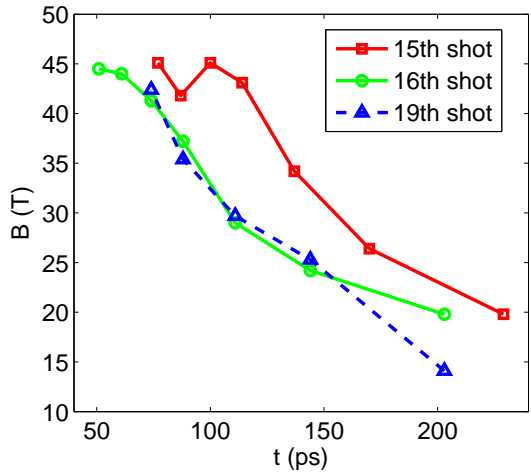}
\caption{(color online) The magnetic field in the center of the circle part calculated by comparing the proton deflection in the straight line part between simulations and experiments is shown.\label{FIG.results}}
\end{figure}

The magnetic field is calculated by comparing the proton deflection in the straight line part between simulations ($l_2$ shown in Fig.\ref{FIG.simulation}(b)) and experiments ($l_1$ shown in Fig.\ref{FIG.simulation}(c) and (d)). The temporal evolution of magnetic field in the center of the circle part is presented in Fig.\ \ref{FIG.results}. It is shown that the magnetic field generating by femtosecond laser beam reaches its maximum in 50-60ps, which is in accordance with the estimation before in this paper. The magnetic field needs more than 100ps to decrease to half of its maximum magnitude, which is long enough to be used in fast electron collimation and other applications. The long decay time of magnetic field is supposed to be caused by the increasing of U-turn coil resistance $R$. When skin effect is considered, $R\sim1\ \Omega$ in room temperature. After the U-turn coil is heated by the current, $R$ could become 10 times larger, and the magnetic field decay time would be $L/R\sim100$ ps ($L\sim10^{-9}$ H)\cite{Daido,Courtois,Fujioka2013}.

The field energy is calculated by integrating the magnetic field energy ($B^2/(2\mu_0)$) in the calculation space ($4\times4\times20$ mm$^3$). The maximum magnetic field obtained in 15th shot is 45T, which corresponds to 40 kA current in the coil, and the field energy calculated is $E_B=1.6$J. The femtosecond laser energy in that shot is 11.1J. So the energy conversion efficiency from laser to magnetic field is 14\%. In this way, the energy conversion efficiency is 22\% and 17\% in 16th shot and 19th shot, respectively. The high efficiency is contributed to the electric potential, which can be developed by the fast electrons escaped from the front and lateral of 1st Ni capacitance disk.


The magnetic field could be optimized by changing the U-turn coil shape according to the applications, such as altering the size of the coil. When 200J picosecond laser is used to produce magnetic field, and assume the energy conversion efficiency from laser to magnetic field is 20\%. As a result, magnetic field with 40J energy could be produced. If the capacity-coil target in our experiments is used (the radius in circle part $R=500\ \mu$m, distance between the straight lines $d=500\ \mu$m), the magnetic field in the center of coil is 227T. If the coil is changed to be smaller, like $R=250\ \mu$m and $d=300\ \mu$m, the center magnetic field would increase to be 595T. But the space region of the magnetic field will be reduced with smaller U-turn coil. It is suggested that magnetic field with amplitude up to 200T would be produced by picosecond laser, and it may be large enough to collimate the electron beam in fast ignition scheme\ \cite{Cai2013}.

In summary, ultra-intense ultra-short laser (femtosecond laser beam) is firstly used to irradiate the capacity-coil target to generate magnetic field. The temporal evolution of magnetic field is observed by proton radiography method. Particle track simulations are used to measure the magnetic field by comparing them with the RCF images. More than 40T magnetic field is obtained by $\sim$10J femtosecond laser beam in the experiments. It makes sure that large magnetic field can also be produced by ultra-intense laser. Although the laser pulse duration is only 30fs, the generated magnetic field can last for 100ps due to the increasing of resistance in U-turn coil. And the energy conversion efficiency from laser to magnetic field can reach as high as $\sim$20\%. The proton hollow structure is explained by 2D3V PIC simulations, which indicate that the electrons in proton beam may have great influence on the proton deflections in Proton Radiography. Finally, the experiments show that tens of tesla(T) magnetic field could be produced in many ultra intense laser facilities around the world. It opens new areas of research that need large magnetic field environment, such as magnetic field reconnection and collisionless shock \cite{Astrophys}.

We gratefully acknowledge Q. Jia, K. A. Tanaka, L. H. Cao, B. F. Shen, C. T. Zhou, H. B. Zhuo for fruitful discussions. This work was supported by the National Nature Science Foundation of China (Grant Nos. 11174259 and 11375160), the Foundation of CAEP (Grant No. 2014A0102003).



\end{document}